\documentclass[prc,twocolumn,showpacs,amsmath,amssymb]{revtex4}
\usepackage{graphicx}
\usepackage{dcolumn}
\usepackage{bm}
\usepackage{longtable}
\usepackage{tabularx}
\newcommand{\bra}[1]{\langle {#1} |}
\newcommand{\ket}[1]{| {#1} \rangle}

\newcommand{\vecr}{{\mathbf r}}

\begin{document}

\title{
Linear response theory in the continuum for deformed nuclei:\\
Green's function vs. time-dependent Hartree-Fock
with the absorbing-boundary condition
}

\author{Takashi Nakatsukasa}
\author{Kazuhiro Yabana}
\affiliation{Center for Computational Sciences and
Institute of Physics, University of Tsukuba, 
Tsukuba 305-8571, Japan}

\date{\today}

\begin{abstract}
The continuum random-phase approximation is extended to
the one applicable to deformed nuclei.
We propose two different approaches.
One is based on the use of
the three dimensional (3D) Green's function and
the other is the small-amplitude TDHF with the absorbing-boundary condition.
Both methods are based on the 3D Cartesian grid representation and
applicable to systems without any symmetry on nuclear shape.
The accuracy and identity of these two methods are examined with
the BKN interaction.
Using the full Skyrme energy functional in the small-amplitude TDHF approach,
we study the isovector giant dipole states in the continuum for $^{16}$O and
for even-even Be isotopes.
\end{abstract}

\maketitle

\section{Introduction}

Mean-field theories with effective interactions
\cite{VB72,Vau73,Wal74,DG80}
have been extensively used for 
systematic description of nuclear ground-state properties 
from light to heavy nuclei, including infinite nuclear matter.
Nuclear mass, radius, density distribution, and deformation are
the primary target of the static effective mean-field theory \cite{LPT03}.
The concept of the nuclear mean-field theory is rather different from
the Hartree-Fock theory in electronic systems but is more close to
the density functional theory.
Especially, the Hartree-Fock (HF) with
the zero-range Skyrme interaction
results in an energy functional of local densities.
A similar form of functional was obtained from the density-matrix
expansion of energy functionals calculated with the
microscopic nucleon-nucleon forces \cite{NV72,NV75}.

Although the static mean-field calculations well reproduce the bulk nuclear
properties throughout the nuclear chart,
it is necessary to go beyond the mean field to describe excited states
and correlations associated with many kinds of collective motions.
The generator coordinate method (GCM) \cite{HW53, GW57} is one of the
standard methods to take account of the configuration mixing.
The GCM based on the mean-field theory provides
a unified description of single-particle
and collective nuclear dynamics.
In practice, collective variables, $q$, are chosen from physical intuition and
are restricted to one dimension in most cases.
For instance, in order to describe quadrupole excitations,
the most common choice is the mass quadrupole moment,
$q=A\bra{\Phi(q)} r^2Y_{20} \ket{\Phi(q)}$, where
the single-Slater states $\ket{\Phi(q)}$
are determined by the constrained Hartree-Fock(-Bogoliubov) calculation.
This is a drawback of the GCM that one has to prepare, {\it a priori},
a set of states $\{\ket{\Phi(q)}\}$


The time-dependent Hartree-Fock (TDHF) theory is a complementary method to
the GCM.   
The system determines its collective path for itself
and the TDHF takes care of both collective and single-particle excitations.
The TDHF is also known to produce the proper inertial parameters \cite{TV62},
because it is a dynamical theory
to incorporate time-odd components in the wave function.
A drawback is its semiclassical nature.
Namely, in order to calculate quantal quantities,
such as eigenenergy and transition probability,
one has to requantize obtained TDHF dynamics.
Although it is a difficult task to requantize the TDHF orbitals
in general \cite{Neg82},
the perturbative regime can be easily handled.
The linear approximation leads to the random-phase approximation (RPA)
for the effective density-dependent forces,
which is analogous to the time-dependent local-density approximation
in electronic systems \cite{RG84,GDP96}.
Another advantage of TDHF is its ability of describing spreading width of
collective motion
induced by the interaction between particles and
time-dependent mean-field potential (one-body dissipation).
The escape width can be also described by the TDHF
but requires proper treatment of the continuum.
In this paper, we propose a feasible method
to treat the continuum in the real-space TDHF calculation.
That is the absorbing-boundary condition (ABC) approach.
We have already studied photoabsorption in molecules \cite{NY01} and
nuclear breakup reaction \cite{UYN02,UYN04} with the similar technique.
Our earlier attempts for nuclear response calculation
have been reported in Refs.~\cite{NY02-P2,NUY03-P,NY04-P1,NUY04-P}.

The Green's function method in the linear response proposed by
Shlomo and Bertsch \cite{SB75} is a common way to treat
the continuum boundary condition.
It is usually called ``continuum RPA'' in nuclear physics.
The same idea was proposed later in the time-dependent density functional
theory (TDDFT) for calculations of photoresponse in rare-gas atoms \cite{ZS80}.
The method has been widely applied to spherical (magic or semi-magic) nuclei
\cite{LG76,TMS80,HS96,HSZ97,HSZ98,HS99,Sag01,SS02,ASS03,AS04}, however,
its application to deformed systems has not been done so far,
because the explicit construction of Green's function 
is extremely difficult for deformed potential.
We have recently proposed an iterative method to construct
response functions for deformed systems with the proper boundary condition
in the three-dimensional (3D) coordinate space
and studied molecular photoabsorption using the TDDFT \cite{NY01,NY03}.
There, the dynamical screening effect in the continuum for a multi-center
problem was a key issue for understanding
the photoabsorption cross section
at photon energies higher than the ionization potential.
The same method is applicable to nuclear mean-field models
that do not contain non-local densities.
In this respect, applications to the Skyrme energy functional is parallel
to the TDDFT.
In this paper, we extend a method of the continuum RPA
to the one in the 3D coordinate space
and apply it to deformed nuclei.
We call this ``3D continuum RPA'' in this paper.
This provides the exact treatment of the nucleonic continuum
for deformed nuclei \cite{NY01-P2}.
The results can be used to check validity of the ABC approach.

Another issue addressed in this paper is
the self-consistent treatment in the continuum response calculation.
The nuclear energy functional is far more complicated than
that of electronic TDDFT.
The Skyrme functional is one of the simplest,
since its non-local part is expressed by derivatives of local densities.
Even so, the continuum RPA calculations so far
neglect the spin-orbit and Coulomb residual particle-hole interactions,
which violates the self-consistency with the HF field \cite{BT75,Sag01}.
In addition, a time-reversal-odd (time-odd) part of densities,
such as spin densities, are often omitted.
Since the spin-orbit term in the time-even mean field
is related to spin-current terms in the time-odd mean field
by the local gauge (Galilean) invariance \cite{Eng75,DD95},
the neglect of spin density violates this symmetry.
As far as we know, at present,
there is no fully self-consistent Skyrme-HF-based continuum RPA
calculations, even for spherical nuclei.
We perform the small-amplitude TDHF calculation with the ABC (TDHF+ABC)
in fully self-consistent manner for the giant dipole resonance in $^{16}$O,
and examine effect of residual interactions which have been neglected so far.
In the time-dependent relativistic mean-field approach without the continuum,
the small-amplitude real-time calculation has been attempted
for spherical nuclei \cite{BVR95,Ring96}.
However, only a very short time period ($3\sim4$ $\hbar$MeV$^{-1}$)
was achieved, which prevents them from carrying out a quantitative analysis.
See also recent papers~\cite{PRNV03,Gia03,Ter04} and references therein
for the present status of
the self-consistent HF(B)+(Q)RPA calculations for spherical nuclei.
It should be noted that, without the continuum boundary condition,
there exist a few works of fully self-consistent RPA
for deformed nuclei calculated in the 3D coordinate space
with the full Skyrme interaction \cite{Ima03,Ina04}.


In recent developments of radioactive-ion-beam facilities,
the Coulomb excitation and the inelastic scattering
are becoming standard methods to investigate excited states in unstable
nuclei.
For weakly bound systems,
the treatment of the continuum should be
extremely important.
Moreover, we know most of open-shell nuclei are deformed.
Collective modes of excitation in the particle continuum in deformed nuclei
become the main interest in those studies.
This has not been examined in a self-consistent manner so far.
The present paper will provide us with general methods of linear response
in the continuum for systems whose energy functional is given by
local one-body densities.

The paper is organized as follows:
Section \ref{sec: 3D_CRPA} presents a method of
extending the continuum RPA to deformed nuclei.
In Sec.~\ref{sec: Real-time}, we present a real-time TDHF
method using the absorbing boundary condition.
Some illustrative examples show
effect of the continuum and
comparison between these two methods in Sec.~\ref{sec: Illustration}.
In Sec.~\ref{sec: Skyrme}, we present
numerical results of the small-amplitude TDHF+ABC calculation in
real time using the Skyrme energy functional for giant dipole resonances.
Effects of time-odd densities, $E1$ strengths in $^{16}$O, and
those in neutron-rich Be isotopes are discussed.
The conclusion is summarized in Sec.~\ref{sec: Conclusion}.

\section{3D Continuum RPA}
\label{sec: 3D_CRPA}

For spherical systems, the continuum RPA is formulated in terms of
the radial Green's function using a multipole expansion \cite{SB75}.
Hereafter, we refer to this as ``1D continuum RPA''.
In Ref.~\cite{NY01}, we have presented a method to construct 
a Green's function in the 3D grid representation for a system 
without any spatial symmetry.
In this section, we recapitulate the method of constructing the
response function.
Spin and isospin indices are suppressed for simplicity
and $\hbar=1$ is used.

The HF Hamiltonian, $h[\rho]$, is a functional of one-body density
matrix \cite{RS80}.
In case of zero-range effective interactions,
it is a functional of local one-body density $\rho(\vecr)$.
The stationary condition is
\begin{equation}
\label{HF}
[h,\rho] = 0,
\end{equation}
which defines the HF ground state density $\rho=\rho_0$.
Then, the TDHF equation with an external perturbation,
\begin{equation}
\label{TDHF}
i\frac{d}{dt}\rho(t) = [h[\rho]+V_{\rm ext}(t),\rho(t)] ,
\end{equation}
is linearized with respect to the density fluctuation,
\begin{equation}
\label{linear_approx}
\rho(\vecr,t)=\rho_0(\vecr)+\delta\rho(\vecr,t) .
\end{equation}
This leads to the well-known RPA equation.
The transition density, $\delta\rho(\vecr;\omega)$,
which is the Fourier transform of $\delta\rho(\vecr,t)$,
can be expressed as \cite{BT75,ZS80}
\begin{eqnarray}
\label{delta_rho_1}
\delta \rho(\vecr;\omega)&=&
                \int\int d\vecr
                \Pi(\vecr,\vecr';\omega) V_{\rm ext}(\vecr';\omega) , \\
\label{delta_rho_2}
            &=&\int d\vecr'
                \Pi_0(\vecr,\vecr';\omega) \nonumber\\
            &\times&\left( V_{\rm ext}(\vecr';\omega)
 +\int d\vecr^{''} v(\vecr',\vecr^{''}) \delta\rho(\vecr^{''};\omega)\right) ,
\end{eqnarray}
where $\Pi$ and $\Pi_0$ are the RPA and the independent-particle response
function, respectively.
The $v(\vecr,\vecr')$ is a residual interaction which is defined by
\begin{equation}
\label{residual_int}
v(\vecr,\vecr')\equiv
\frac{\delta^2 E[\rho]}{\delta\rho(\vecr)\delta\rho(\vecr')} .
\end{equation}
Here, we assume that the total energy functional, $E[\rho]$, is
a function of local density $\rho(\vecr)$ only.
The HF mean field is also local in the coordinate space.
Assuming that a one-particle moment $F(\vecr)$ depends only on the spatial
coordinates,
the transition strength is obtained from the transition density,
\begin{eqnarray}
\frac{dB(\omega; F)}{d\omega} &\equiv&
 \sum_n |\bra{n}F\ket{0}|^2 \delta(\omega-E_n) ,\\
 &=& -\frac{1}{\pi} \mbox{Im} \int\int d\vecr d\vecr'
      F(\vecr) \Pi(\vecr,\vecr';\omega) F(\vecr') ,\\
 &=& -\frac{1}{\pi} \left. \mbox{Im} \int d\vecr
      F(\vecr) \delta\rho(\vecr;\omega)\right|_{V_{\rm ext}=F} .
\label{BF_Green}
\end{eqnarray}
In case of deformed nuclei,
$\ket{0}$ and $\ket{n}$ are not eigenstates of total angular momentum
operator.
Thus, $dB(\omega; F)/d\omega$ should be regarded as
the intrinsic transition strength.
In Sec.~\ref{sec: Be}, we assume the strong coupling scheme \cite{BM75}
in order to transform calculated intrinsic strength to the quantity in
the laboratory frame.
The response function is written as
\begin{eqnarray}
\label{Pi}
&&\Pi(\vecr,\vecr';\omega)=\Pi_0(\vecr,\vecr';\omega) \nonumber \\
  &+&\int\int d\vecr^{''} d\vecr^{'''}
     \Pi_0(\vecr,\vecr^{''};\omega) v(\vecr^{''},\vecr^{'''})
                        \Pi(\vecr^{'''},\vecr';\omega) ,\\
\label{Pi_0}
&& \Pi_0(\vecr,\vecr';\omega) =
 \sum_{i=1}^A \left\{
     \phi_i(\vecr)
     G^{(-)}(\vecr,\vecr';\epsilon_i-\omega) \phi_i^*(\vecr')
   \right. \nonumber \\
   && \hspace{50pt}+ \phi_i^*(\vecr)
   \left. G^{(+)}(\vecr,\vecr';\epsilon_i+\omega) \phi_i(\vecr') \right\} .
\end{eqnarray}
The single-particle Green's function in Eq.~(\ref{Pi_0}) is defined by
\begin{equation}
\label{G_sp}
G^{(\pm)}(\vecr,\vecr';E)
  = \bra{\vecr}\left( E-h[\rho_0] \pm i\eta \right)^{-1}\ket{\vecr'} .
\end{equation}
Here, the superscript $+(-)$ indicates
the outgoing (incoming) boundary condition.
In case that $h[\rho_0]$ is rotationally invariant,
the Green's function of Eq.~(\ref{G_sp}) can be constructed by using
the partial-wave expansion,
\begin{equation}
\label{G_pwe}
G^{(\pm)}(\vecr,\vecr';E)= 2m
 \sum_{lm}
  Y_{lm}(\hat\vecr) \frac{u_l(r_<) w_l^{(\pm)}(r_>)}{W[u_l,w_l^{(\pm)}]\ rr'}
  Y_{lm}^*(\hat{\vecr}') .
\end{equation}
Here, $W$ is the Wronskian and
$u_l$ and $w_l$ are solutions of the radial Schr\"odinger equation
for $h[\rho_0]=-\nabla^2/2m + V(r)$:
\begin{equation}
\label{radial_HF_eq}
  \left( E + \frac{1}{2m}\frac{d^2}{dr^2}-\frac{l(l+1)}{2mr^2}-V(r) \right)
  R_l(r)=0 .
\end{equation}
$u_l$ is regular at origin and $w_l^{(\pm)}$ has an outgoing/incoming
asymptotic form.
In the 1D continuum RPA \cite{SB75},
Eq.~(\ref{Pi}) is also expanded in partial waves.
Then, the 1D RPA response function (in the radial coordinate)
is explicitly constructed by using Eqs.~(\ref{Pi}), (\ref{Pi_0}),
and (\ref{G_pwe}).

There are some difficulties to extend the theory to non-spherical systems.
The first one is a purely numerical difficulty.
Since the number of spatial grid points in the 3D space
is much larger than that of the radial grid points, it is hard
to explicitly construct the response function,
$\Pi(\vecr,\vecr';\omega)$ and
to perform spatial multi-fold integration.
We also need to calculate an inverse matrix to solve Eq.~(\ref{Pi}).
The second difficulty lies in the complexity of boundary condition.
Equations (\ref{G_pwe}) and (\ref{radial_HF_eq})
cannot be used for cases of a deformed HF potential.

We solve the first numerical problem by using an iterative procedure for
implicit calculation of the response and Green's function.
For instance, in order to calculate the transition density,
we recast Eq.~(\ref{delta_rho_2}) into an integral equation
for $\delta\rho$,
\begin{eqnarray}
\label{RPA_iteration}
&&\int d\vecr^{''} \left\{ \delta(\vecr-\vecr^{''})
                   - \int d\vecr' \Pi_0(\vecr,\vecr';\omega)
         v(\vecr',\vecr^{''}) \right\} \delta\rho(\vecr^{''};\omega)
  \nonumber \\
&&= \int d^3 r' \Pi_0(\vecr,\vecr';\omega) V_{\rm ext}(\vecr';\omega) .
\end{eqnarray}
This is equivalent to a linear algebraic equation in the 3D grid space
and we use the iterative method to solve it.
For the linear algebraic problem, $A\ket{x}=\ket{b}$,
the iterative methods require neither a full knowledge of the matrix $A$
nor an inverse matrix $A^{-1}$,
but do only results of operating $A$ on a certain vector $\ket{y}$.
This means that we do not need to calculate an explicit form of
$\Pi_0(\vecr,\vecr';\omega)$.
All we need to calculate is the action of $\Pi_0$; i.e.,
$\Pi_0 \cdot v \cdot \delta\rho$ and $\Pi_0\cdot V_{\rm ext}$
where the dot indicates the integral in Eq. (\ref{RPA_iteration}).
This is an advantage of the iterative method over the direct method.
In addition, the iterative method is known to be very efficient
for a large sparse matrix.
Then, the next task is to calculate action of $\Pi_0$.
According to Eq. (\ref{Pi_0}),
we have to operate $G^{\pm}(E)$ on certain states $\ket{y}$.
Now the problem is coupled to the second difficulty, that is
the continuum boundary condition for deformed systems.

We start to divide the HF potential into a long-range spherical part
and a short-range deformed one,
$h[\rho_0]=-\nabla^2/2m+V_0(r)+\tilde{V}(\vecr)$.
In the present work, $V_0(r)$ is taken as the Coulomb potential of
a sphere of radius $1.2A^{1/3}$ fm with a uniform change $Ze$.
The single-particle Green's function
for $h_0=-\nabla^2/2m+V_0(r)$ is constructed in the same way as
Eq. (\ref{G_pwe}) which is denoted by $G_0^{(\pm)}(E)$ below.
We have an identity for $G$,
\begin{eqnarray}
\label{Dyson_eq}
&& G^{(\pm)}(\vecr,\vecr';E) = G_0^{(\pm)}(\vecr,\vecr';E)\quad\quad \nonumber \\
&&\hspace{5pt}+ \int d^3r^{''}
 G_0^{(\pm)}(\vecr,\vecr^{''};E)
               \tilde{V}(\vecr^{''})
               G^{(\pm)}(\vecr^{''},\vecr';E) .
\end{eqnarray}
The boundary condition of $G_0^{(\pm)}$ determines an asymptotic
behavior of $G^{(\pm)}$.
The action of $G^{(\pm)}$,
$\ket{x^{(\pm)}}=G^{(\pm)}\ket{y}$ for a given state $\ket{y}$,
is obtained by solving a linear algebraic equation
\begin{equation}
\label{G_iteration}
\left\{ 1 - G_0^{(\pm)}\tilde{V}\right\} \ket{x^{(\pm)}}=G_0^{(\pm)}\ket{y} .
\end{equation}
Here, we use, again, the iterative method to solve this equation.

In summary, to obtain the transition density, we solve
Eq.~(\ref{RPA_iteration}).
In order to do this,
we need to calculate the operation of $\Pi_0$, which then requires
us to solve Eq.~(\ref{G_iteration}) with a proper boundary condition.
The procedure results in multiple-nested linear algebraic equations
which are solved with iterative methods, such as the conjugate gradient
method.
The detailed algorithm is given in Ref.~\cite{NY01}.

\section{Real-time TDHF+ABC}
\label{sec: Real-time}

\subsection{Absorbing boundary condition (ABC)}
The TDHF equation can be efficiently solved in
the 3D lattice space in real time \cite{FKW78,BGK78,Neg82}.
The same technique has been applied to TDDFT of finite \cite{YB96,YB99}
and infinite electronic systems \cite{BIRY00}.
In the real-time calculation, we propagate single-particle wave functions
$\{\phi_i\}_{i=1,\cdots,A}$ using the same technique as that
in Ref.~\cite{FKW78}.
\begin{equation}
\label{time_evolution}
\phi_i(\vecr,t+\varDelta t)
  =\exp(-i\varDelta t\cdot h[\rho(t+\varDelta t/2)])\phi_i(\vecr,t) ,
\end{equation}
where the exponential operator is expanded in a power series to
$(\varDelta t)^4$.
The time step in following applications is taken as
$\varDelta t=0.001$ MeV$^{-1}$.
There are many good reasons for solving the problem
in real-time representation.
First, the computation algorithm becomes very simple.
In order to make the time evolution,
only the operation of the HF Hamiltonian on a certain single-particle state,
$h[\rho]\ket{\psi}$, is needed to be calculated,
though the self-consistency between $\rho(t)$ and $h[\rho]$ brings slight
complication.
Secondly, a single calculation of the time evolution
provides information for a wide range of energy.
Thus, for the strength function in a wide energy region,
the real-time calculation is often more efficient than
that in the energy domain.
Last but not least, the TDHF wave packet in real space in real time
gives an intuitive understanding of dynamics.
In pioneering works on heavy-ion collisions,
the TDHF dynamics nicely demonstrated time evolution of
nuclear inelastic scattering \cite{FKW78,CMM78,Neg82}.

The TDHF time evolution is relatively easy in the present computer power.
The problem is how to impose the continuum boundary condition.
The exact treatment of the continuum such as the Green's function method
is very difficult (even impossible) in this case,
because the energy of escaping particles cannot be determined uniquely.
Thus, we attempt an approximate treatment.
The usual approach to TDHF in real space is to assume the wave function
to be zero at some distance $R$ from the origin,
which we call ``Box boundary condition (BBC)'' hereafter.
Then, the time evolution must be completed before a significant portion of
the wave reaches the boundary.
Seeking higher accuracy, we must employ a larger $R$ value,
increasing the computation task.
Instead, in this paper, we employ the ``Absorbing boundary condition (ABC)'',
which introduces a complex absorbing potential outside of the system.
The method was first tested by Hamamoto and Mottelson
for a schematic one-dimensional model of TDHF calculation \cite{HM77}.
Since then, however, its capability has not been fully examined
in nuclear theory.
On the other hand, in other fields of quantum physics,
especially in atomic and molecular collision theories,
the method has become one of standard methods for calculations of
reactive scattering problems
(see a recent review article \cite{MPNE04} and references therein).
We have demonstrated that the ABC is able to produce
results identical to that of the exact continuum with
the Green's function for
TDDFT study of photoabsorption in molecules \cite{NY01}.
In nuclear three-body reaction models \cite{UYN02},
the method was also tested in detail for deuteron breakup reaction
and provides an alternative method to the
continuum-discretized-coupled channels (CDCC).
A similar approach has been tested to calculate nuclear resonance states in
a simple model \cite{MH02}.

The success of the ABC approach is based on its simplicity.
Actually, it requires only a minor modification of the real-time
TDHF code, simply adding a complex potential, $-i\tilde\eta(\vecr)$.
We replace the HF Hamiltonian in Eq.~(\ref{time_evolution}) by
\begin{equation}
h[\rho] \longrightarrow h[\rho]-i\tilde\eta(\vecr) .
\end{equation}
This prescription is equivalent to the use of Green's function
of Eq.~(\ref{G_sp}) in which the infinitesimal imaginary part
$\eta$ is replaced by a finite and coordinate-dependent $\tilde\eta(\vecr)$.
The absorbing potential must be zero in a region where
the ground-state density has a finite value,
and it is finite outside of the system.
Of course, the addition of the complex potential violates
the unitarity of time evolution.
Thus, the norm of each single-particle state decreases with time,
which represents a physical process, the emission of particles.

We adopt the same form of absorptive potential as previous works
\cite{NY01,UYN02}.
This is a linear dependence on the coordinate \cite{NB89,Chi91}:
\begin{equation}
\label{absorbing_pot}
\tilde\eta(r)=\left\{
\begin{array}{ll}
0 & \mbox{ for } 0< r < R ,\\
i\eta_0\frac{r-R}{\varDelta r}  & \mbox{ for }  R < r < R + \Delta r.
\end{array}
\right.
\end{equation}
The size of the inner model space ($r<R$) is chosen so that
the HF ground state converges within this space.
The outer space of width $\varDelta r$ ($R<r<R+\varDelta r$) is
the absorbing region that should be large enough to prevent
reflection of emitted outgoing waves.
The condition of a good absorber for a particle with mass $m$ and energy $E$
is given by
\begin{equation}
\label{absorb_cond}
7 \frac{E^{1/2}}{\Delta r \sqrt{8m}} < | \eta_0 |
 < \frac{1}{10} \Delta r \sqrt{8m} E^{3/2} .
\end{equation}
Here, we demand the reflection smaller than 0.1\% and
the transmission smaller than 3.3\%.
The similar condition was given in Refs.~\cite{NB89,Chi91,NY01}.
Since the condition is energy dependent,
we choose $\varDelta r$ and $W_0$ as,
\begin{equation}
\varDelta r=12 \mbox{ fm,} \quad\quad \eta_0=10\mbox{ MeV}.
\end{equation}
This satisfy the condition of Eq. (\ref{absorb_cond})
for $7<E<60$ MeV.

For the linear response calculation,
first, we solve the static HF problem with
the imaginary-time method \cite{DFKW80}
to determine the occupied HF orbitals $\{\phi_i^0\}_{i=1,\cdots,A}$.
Then, an external perturbative field,
$V_{\rm ext}(\vecr,t)=\epsilon F(\vecr) \delta(t)$,
is turned on instantaneously at $t=0$.
This results in an initial state of the TDHF calculation as
\begin{equation}
\phi_i(\vecr,t=0+)=e^{-i \epsilon F(\vecr)} \phi_i^0(\vecr) ,
\end{equation}
where the constant $\epsilon$ is arbitrary but should be small enough to
validate the linear approximation of Eq. (\ref{linear_approx}).
We calculate time evolution of the expectation value of $F(\vecr)$
(assumed to be real),
\begin{eqnarray}
\bra{\Psi(t)}F\ket{\Psi(t)} &=& \int d\vecr \sum_{i=1}^A
         \phi_i^*(\vecr,t) F(\vecr) \phi_i(\vecr,t) \nonumber \\
    &=& \int d\vecr F(\vecr) \delta\rho(\vecr,t) .
\end{eqnarray}
Here, we assume that the ground-state expectation value of $F(\vecr)$
is zero at the last equation.
Comparing its Fourier transform
with Eq. (\ref{BF_Green}), we have
\begin{equation}
\label{BF_RT}
\frac{dB(\omega; F)}{d\omega}=
-\frac{1}{\pi\epsilon}\mbox{Im} \int dt \bra{\Psi(t)}F\ket{\Psi(t)}
 e^{i\omega t} .
\end{equation}
Note that $\bra{\Psi(t)}F\ket{\Psi(t)}$
is fully determined by wave functions in 
the inner space ($r<R$), as far as the linear approximation is valid.
This can be easily understood using a relation,
$\delta\rho=\sum_i \phi_i^{0*} \delta\phi_i + \mbox{ h.c.}$,
and the condition that $\phi_i^0=0$ at $r>R$.

\subsection{Adaptive-coordinate 3D grid space}
\begin{figure}[th]
\includegraphics[width=0.4\textwidth]{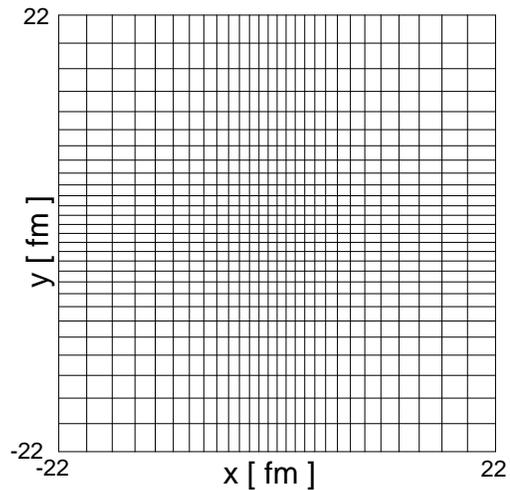}
\caption{\label{fig: grid}
Adaptive grid in the $(x,y)$-plane for
the coordinate transformation of Eq. (\ref{adaptive}) with
$x_0=8$ fm, $k=5$ and $n=2$.
The $(u,v,w)$-space is discretized in square mesh of 0.9 fm.
}
\end{figure}

There is a significant improvement of the computational cost
by reducing the number of grid points in the outer space.
Although we take the outer model space roughly the same size as the
inner one in the radial coordinate ($R\approx \varDelta r$),
its volume is considerably larger
because the volume element increases as $r^2 dr$.
Therefore, the calculation of wave functions in the outer model space
consumes most part of the computation time.
However, since wave functions in the outer space is irrelevant
for $\bra{\Psi(t)}F\ket{\Psi(t)}$,
the accurate description is not necessary there.
Therefore, we use the adaptive curvilinear coordinate
in the small-amplitude TDHF+ABC calculation \cite{MZK97}.
The coordinate transformation we use in this paper is
\begin{equation}
\label{adaptive}
x(u)=x_0 \frac{ku/x_0}{1+(k-1)u/(x_0 \sinh(u/x_0))^n},
\end{equation}
and the same form for $y(v)$ and $z(w)$ as well.
This function has an asymptotic values, $x(u)\sim u$ at $u\ll x_0$ and
$x(u)\sim ku$ at $u\gg x_0$.
All the derivatives and integrals in $(x,y,z)$-space are
mapped to those in $(u,v,w)$-space.
For instance,
\begin{eqnarray}
\frac{\partial^2}{\partial x^2} &=&
              \frac{d^2u}{dx^2}\frac{\partial}{\partial u} +
            \left( \frac{du}{dx} \right)^2 \frac{\partial^2}{\partial u^2} ,\\
\int d\vecr &=& \int dudvdw \frac{dx}{du} \frac{dy}{dv} \frac{dz}{dw} .
\end{eqnarray}
The 3D $(u,v,w)$-space is discretized in square mesh
and finite-point formula in this space is applied to numerical differentiation.
The curvilinear grid space employed in Secs.~\ref{sec: Illustration} and
\ref{sec: Skyrme} is shown in Fig.~\ref{fig: grid}.

\section{Illustrative examples: Application with the BKN interaction}
\label{sec: Illustration}

In this section, some illustrative examples are shown to demonstrate
effects of continuum, validity of bound-state ($L^2$) approximation,
and comparison between Green's function and ABC approach.
We adopt the BKN interaction used in Ref.~\cite{FKW78}.
Note that, for this schematic interaction,
the spin-isospin degeneracy is assumed all the time and
the Coulomb potential acts on all orbitals with a charge $e/2$.
The HF one-body Hamiltonian is given by
\begin{equation}
\label{BKN}
h[\rho]=-\frac{1}{2m}\nabla^2 + \frac{3}{4}t_0\rho + \frac{3}{16}t_3 \rho^2
+W_Y + W_C ,
\end{equation}
where the Yukawa potential, $W_Y$, and Coulomb potential, $W_C$,
consist of their direct terms only.
The parameters are taken from Table. I of Ref.~\cite{FKW78}.

\begin{figure}[th]
\includegraphics[width=0.45\textwidth]{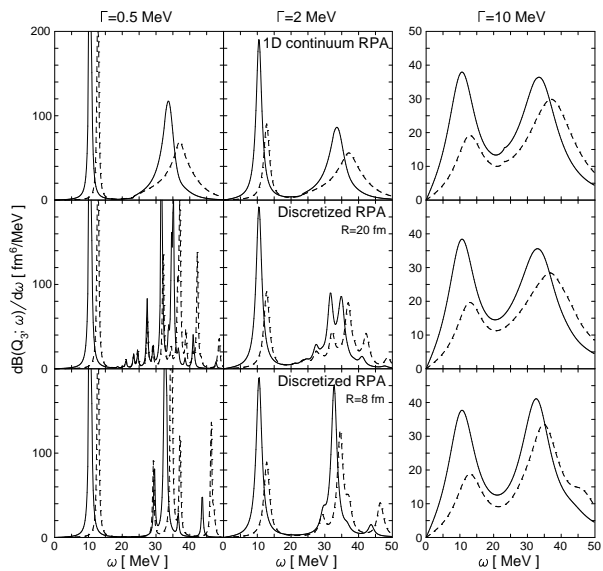}
\caption{\label{fig: O16_oct}
Mass octupole strengths as functions of excitation energy
for $^{16}$O calculated with the BKN interaction.
The top panels show results of the self-consistent continuum RPA.
The middles and bottoms show results of the discretized RPA of
$R=20$ and 8 fm, respectively.
The smoothing parameter $\Gamma$ increases from left to right,
0.5, 2, and 10 MeV.
The solid curves show the RPA strength function, while the dashed
show unperturbed one.
Since each orbital has a four-fold degeneracy (spin-isospin),
the $E3$ strengths are those shown multiplied by $e^2/4$.
}
\end{figure}

\subsection{Continuum in the spherical nucleus: $^{16}$O}
\label{sec: O16}
\subsubsection{$L^2$-approximation of the continuum}
\label{sec: L2-approx}

RPA calculations are often performed on the $L^2$ basis set,
such as the harmonic oscillator basis.
Bound excited states are well described in those calculations, but
how accurate is the $L^2$-approximation for resonance and continuum states?
In other words, what size of model space is necessary to describe
an excited state with a finite life time?
In Ref.~\cite{NY02-P2}, we give a relation between
the box size $R$ and the energy resolution $\varDelta E$
for continuum states;
$\varDelta E \sim \hbar v/R$,
where $v$ is the velocity of an escaping particle.
If we consider a resonance with life-time of $\tau$,
we should read $\varDelta E \sim \hbar v/(R+v\tau)$.
For a long-lived state, $v\tau \gg R$,
this is identical to the uncertainty principle,
$\varDelta E\sim \hbar/\tau$.
However, for a state of $v\tau \ll R$, such as broad resonance
and non-resonant continuum,
the resolution is limited by the size of model space.

Using the BKN interaction, it is easy to perform the self-consistent
1D continuum RPA calculation for closed-shell spherical nuclei.
We show, in Fig.~\ref{fig: O16_oct},
results of the 1D continuum RPA and the RPA in a box radius $R$ with BBC,
which is referred to as "discretized RPA",
for isoscalar (IS) octupole resonance in $^{16}$O.
We should note that a similar study on monopole resonance
can be found in Ref.~\cite{NY02-P2}.
For the continuum RPA calculation, the outgoing boundary condition is
imposed at $r=8$ fm (top panels).
For simplicity,
we use the free asymptotic form, $w_l^{(\pm)}\sim e^{\pm ikr}$
with $k^2/2m=\left[ E-V_C(r)-l(l+1)/2mr^2\right]$ at $r=8\mbox{ fm}$,
instead of the exact Coulomb wave function.
For the discretized RPA, the radius of model space is chosen as
$R=20$ fm (middle panels) and $R=8$ fm (bottom).
Since the discretized RPA produces only discrete peaks,
we use a smoothing parameter $\Gamma$, adding
an imaginary part, $i\Gamma/2$, to the real energy $\omega$.
In the continuum calculation, though we do not need to smear out
the continuum strength,
we use the same value of $\Gamma$ to make the resolution
as coarse as the discretized RPA.

The continuum RPA calculation clearly shows
the low-energy (LEOR)
and high-energy octupole resonance (HEOR).
The single-particle energy for $p$-shell is about $-16$ MeV in
this calculation.
Thus, the LEOR is a bound peak whose width is entirely from a
smoothing parameter $\Gamma$.
As you see in Fig.~\ref{fig: O16_oct},
the bound LEOR peak depends neither on the boundary condition nor on
the box size $R$.
This justifies the use of the discretized RPA for bound excited states.
On the other hand, the structure of HEOR,
which is embedded in the continuum, strongly depends on values of $R$.
Note that the continuum RPA results with $\Gamma=0.5$ MeV is almost
identical to that with $\Gamma=0$, which means that the width of HEOR
is not artificial in contrast to the LEOR.
The parameter $\Gamma$ actually controls the energy resolution.
For the discretized calculations with $R=20$ fm,
we need $\Gamma \gtrsim 8$ MeV to produce roughly identical results to
the continuum calculation.
For those with $R=8$ fm,
we still see some discrepancy even with $\Gamma=10$ MeV.
In order to obtain sensible results in the discretized basis,
we should average the strength function
with $\Gamma$ inversely proportional to the box size.
We find an empirical formula,
$\Gamma\approx 3 (\hbar v/R)$ for this calculation.
The continuum results with $\Gamma=0.5$ MeV can be reproduced by
the discretized calculation
if we employ a model space of $R \gtrsim 200$ fm.
It is nearly impossible to treat this size of the 3D grid space
with a present computer power.
Therefore, it is certainly desirable to develop a method
of treating the continuum boundary condition for deformed nuclei.
The methods described in Secs.~\ref{sec: 3D_CRPA} and \ref{sec: Real-time}
will serve this purpose.
Next, we discuss applications of these methods
to the BKN interaction.

\subsubsection{Small-amplitude TDHF+ABC vs. continuum RPA}
\label{sec: Comparison}

\begin{figure}[th]
\includegraphics[width=0.35\textwidth]{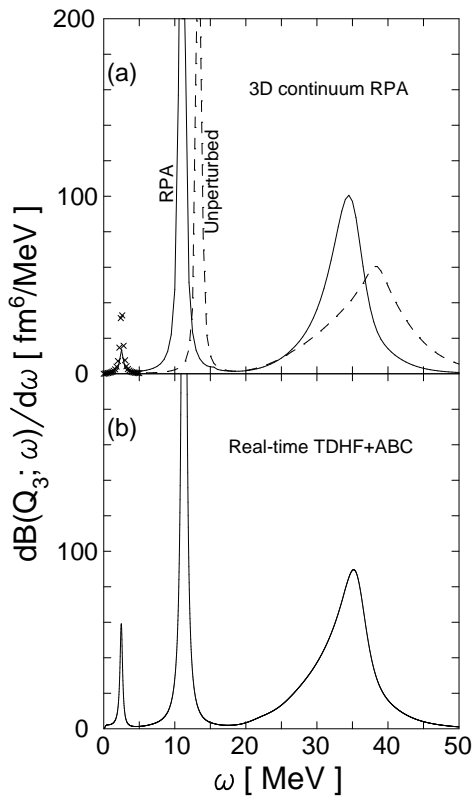}
\caption{\label{fig: 3D_ABC}
The same as Fig.~\ref{fig: O16_oct}, but calculated with
(a) the Green's function method in the 3D grid space
and (b) the real-time small-amplitude TDHF+ABC in the adaptive 3D space.
The smoothing parameter is $\Gamma=0.5$ MeV.
Calculated spurious dipole strength is
shown by crosses for $0<\omega<5$ MeV in units of fm$^2$.
}
\end{figure}
\begin{figure}[th]
\includegraphics[width=0.45\textwidth]{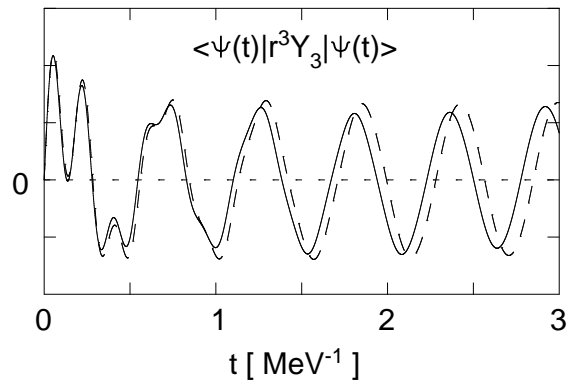}
\caption{\label{fig: RT}
Time evolution of the octupole moment as a function of time
for $^{16}$O.
The calculation in the adaptive (square) mesh coordinate is shown by the
solid (dashed) line.
}
\end{figure}

In this section, we examine accuracy and feasibility of methods
in Sec.~\ref{sec: 3D_CRPA} and in Sec.~\ref{sec: Real-time}.
We compare results of two methods and show how accurate
the TDHF+ABC can be.

We use the same BKN model, (\ref{BKN}),
and, again, calculate octupole states in $^{16}$O.
For the 3D continuum RPA calculation,
the model space is the 3D coordinate space of $R=8$ fm,
discretized in square mesh of
$\varDelta x=\varDelta y=\varDelta z=1$ fm.
For the real-time method with small-amplitude TDHF+ABC,
we use the adaptive curvilinear coordinate of Fig.~\ref{fig: grid},
with $R=8$ fm and $\varDelta r=12$ fm,
and $\eta_0=10$ MeV for the absorbing potential.
The time evolution is calculated up to $t=30$ MeV$^{-1}$, then
we perform the Fourier transform of the time-dependent octupole moment
with $F(\vecr)=r^3Y_{30}(\hat{\vecr})$.
Results of the Green's function method is shown
in Fig.~\ref{fig: 3D_ABC}~(a) and those of the real-time method in
Fig.~\ref{fig: 3D_ABC}~(b).
These figures are almost identical to each other, but one may notice
small difference.
First, the strength at $20<\omega<30$ MeV is slightly higher for the
real-time calculation.
This is probably because the condition for the absorber, (\ref{absorb_cond}),
breaks down for low-energy particles ($E<7$ MeV).
Secondly, the peak position is higher for the real-time calculation by
about 0.3 MeV for LEOR and about 0.6 MeV for HEOR.
This is due to the use of adaptive coordinate representation.
The time evolution of octupole moment is shown in Fig.~\ref{fig: RT}
with use of the square and the adaptive coordinate.
Discrepancy seen at $t>1$ MeV$^{-1}$ corresponds to 0.3 MeV difference
in the LEOR energy.
Comparing results of these calculations with those
of the 1D continuum RPA in the radial coordinate,
we see a very good agreement (see the top-left panel of
Fig.~\ref{fig: O16_oct}).
This means that the nucleonic continuum states are properly treated in
both calculations of the 3D coordinate space;
the Green's function method and the small-amplitude TDHF+ABC.
There is a small peak in the 3D calculation at $\omega=2.5$ MeV.
This is due to small admixture of the spurious translational mode.
In Fig.~\ref{fig: 3D_ABC}~(a), the strength calculated with
$V_{\rm ext}=rY_{10}$ is presented by crosses for $0<\omega<5$ MeV
in units of fm$^2$.
In the 1D continuum RPA calculation with the partial-wave
expansion, these octupole and dipole modes are separated.
Thus, this mixing is not present in Fig.~\ref{fig: O16_oct}.
However, in the 3D grid space,
the translational and rotational invariance of the
Hamiltonian is not exact.
Adopting finer grid spacing diminishes the spurious peak height and
moves its position toward zero energy.

Figure~\ref{fig: RT} demonstrates an interesting feature in real time.
The total energy is conserved within 300 keV up to $t=30$ MeV$^{-1}$.
With the BBC instead of the ABC, the energy conservation becomes even better.
The absolute scale of its vertical axis
does not have a significant meaning because it depends
linearly on the arbitrary small parameter $\epsilon$.
In the beginning, there is interference between the LEOR and HEOR,
however, for $t\gtrsim 1$ MeV$^{-1}$, only the LEOR mode survives.
This feature clearly indicate stability of the bound collective excitation
and decay of the collective mode in the continuum.
The single-mode oscillation of the LEOR continues to the end of
the time evolution ($t=30$ MeV$^{-1}$).
The HEOR decays into the nucleon emission
within time scale of $t \sim 0.5$ $\hbar/$MeV.
Therefore, a part of the calculated energy width of HEOR,
at least a few MeV, is associated with this nucleon escape width.

\begin{figure}[th]
\includegraphics[width=0.4\textwidth]{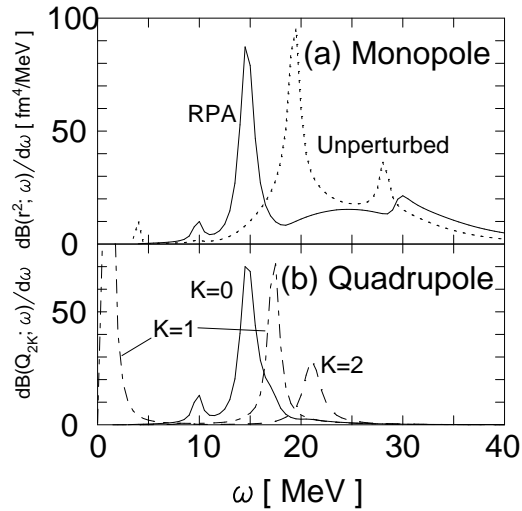}
\caption{\label{fig: Ne20_MQ}
Calculated strength functions for $^{20}$Ne.
The smoothing parameter $\Gamma=1$ MeV is used.
(a) IS monopole resonance.
The solid (dotted) line indicates the RPA (unperturbed) strengths.
(b) IS quadrupole resonance.
The $K=0$, $K=1$, and $K=2$ quadrupole strengths are
shown by solid, dash-dotted, and dashed lines, respectively.
}
\end{figure}

\subsection{Continuum in the deformed nucleus: $^{20}$Ne}
\label{sec: 20Ne}

Now let us discuss a light deformed nucleus, $^{20}$Ne.
This illustrates usefulness and difficulties of the present approaches.
Using the BKN interaction, $^{20}$Ne has a superdeformed prolate shape
with $\beta\approx 0.6$. 
This nucleus has a ground-state rotational band and the measured
$B(E2;\  2^+\rightarrow 0^+)$ value is consistent with the deformation.
Former calculations of the variation after parity projection have produced
the $Y_{30}$-type octupole-deformed ground state with the BKN \cite{TYI96}
and with the Skyrme interaction \cite{OYN04}.
Since the system is deformed,
the 1D continuum RPA is no longer applicable.
This is the first attempt of the 3D continuum RPA calculation for deformed
nuclei.

We use the same model space as the previous calculation on $^{16}$O.
The IS monopole ($r^2$) and quadrupole field ($r^2Y_{2K}(\hat{\vecr})$)
are adopted as the external perturbations in Eq. (\ref{delta_rho_1}).
Results of the 3D continuum RPA are shown in Fig.~\ref{fig: Ne20_MQ}.  
The calculated single-particle energy of the last occupied orbital
is $-10.8$ MeV.
Thus, all the high-energy peaks in the figure are embedded in the continuum.
The giant quadrupole resonance shows three peaks in order of
$K=0$, 1, and 2 in increasing energy (Fig.~\ref{fig: Ne20_MQ} (b)).
Energy spacing between $K=0$ and 1 peaks is smaller than that between
$K=1$ and 2.
This agrees with the simple scaling rule \cite{NA85}.
The result also indicates no low-energy quadrupole vibration
except for the zero-mode with $K=1$.
This is a characteristic feature in the superdeformation \cite{NMM92,NMMS96}.

The monopole strength seems to consist of two components:
a peak at 15 MeV and a broad hump in the energy region of $E>20$.
The peak position is lower than that of the unperturbed peak, by about 5 MeV.
For the monopole strength in $^{16}$O,
calculated strength is shifted higher in energy with the BKN interaction
\cite{NY02-P2}.
Therefore, we consider this lowering in energy due to
strong coupling to the quadrupole resonance.
In fact, the peak lies at exactly the same energy as the $K=0$ quadrupole
resonance (Fig.~\ref{fig: Ne20_MQ}~(b)).
We have reported a similar result for the oblate nucleus, $^{12}$C
\cite{NY02-P1}.
Although the BKN interaction may not be realistic for arguing real phenomena in
$^{20}$Ne, effects of such coupling in the continuum between
different multipole resonances in deformed nuclei would be an interesting
subject in future.
There are experimental data on this issue \cite{Gar80,YLC99,Ito01}.

At the end of this section,
we would like to mention a numerical problem of the real-time TDHF+ABC method.
We have a difficulty to calculate a certain class of IS modes of excitation
with the real-time method.
This is associated with zero (Nambu-Goldstone) modes.
For instance, calculating TDHF time evolution with the external perturbative
IS $K=0$/$K=1$ octupole field,
the center of mass of the nucleus starts moving because of coupling to
the translational motion.
Of course, if we adopt a very small grid spacing, these modes are
decoupled, which is guaranteed by the self-consistent HF+RPA theory.
In practice, we use a mesh of order of 1 fm in the 3D Cartesian coordinates
and a finer mesh size drastically increases a computational task.
The problem is more serious in deformed cases than in spherical,
because the angular momentum selection rule no longer works.
In addition, the deformed nucleus has the rotational mode as another
zero mode which is clearly seen in  the $K=1$ mode in Fig.~\ref{fig: Ne20_MQ}.
In $^{16}$O, we are able to perform the time evolution up to
$t\geq 30$ MeV$^{-1}$, however, 
for the $K=0$ octupole mode in $^{20}$Ne , $t\approx 10$ MeV$^{-1}$ is
a limit of time period in which the reliable calculation can be done.
This is, of course, a matter of computational cost.
If we do not use the adaptive curvilinear coordinate and adopt
a larger space, we can carry out a stable calculation for a longer period.
Because of this problem,
we shall discuss applications with the Skyrme interaction to
the isovector (IV) giant dipole resonances (GDR) in the next section,
which is more stable and feasible.

\section{GDR studied with Skyrme TDHF+ABC}
\label{sec: Skyrme}

\subsection{Effects of time-odd mean field in $^{16}$O}
\label{sec: 16O}

\begin{figure}[ht]
\includegraphics[width=0.42\textwidth]{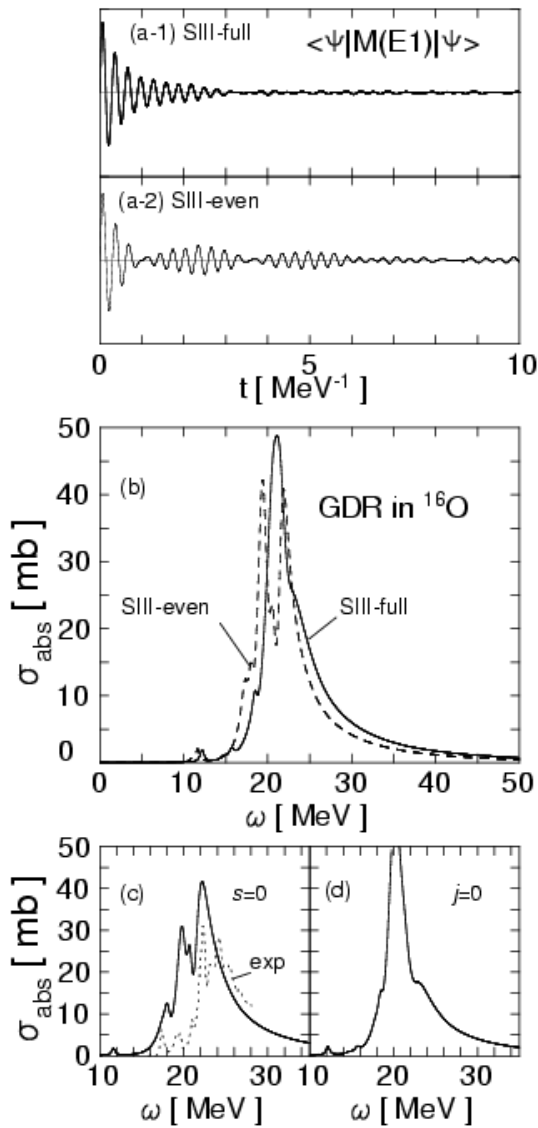}
\caption{\label{fig: O16_GDR}
Results of the Skyrme TDHF+ABC for GDR in $^{16}$O.
(a-1) Time evolution of the $E1$  moment as a function of time
calculated with the SIII-full.
(a-2) The same as (a-1) but with SIII-even.
(b) Calculated photoabsorption cross section
as a function of excitation energy.
The SIII-full calculation (solid line) is
compared to the SIII-even (dashed line).
The smoothing parameter $\Gamma=0.5$ MeV is used.
(c) The same as (b) but neglecting the spin density $\mathbf{s}$.
Experimental photoneutron cross section ($\times 3.5$) is shown by the
dotted line \cite{BF75}.
(d) The same as (b) but neglecting the current density $\mathbf{j}$.
See text for details.
}
\end{figure}

The continuum RPA with the Skyrme energy functional
is a standard method for describing collective excitations
in closed-shell spherical nuclei.
However, its fully-self-consistent calculations have not been achieved
in practice, neglecting residual spin-orbit and Coulomb interactions.
In addition, some of the time-odd densities, which are known to be important
for nuclear moment of inertia and the local Galilean invariance \cite{DD95},
are often neglected in the continuum RPA.
In this section, we present an application of
the small-amplitude TDHF+ABC method to the giant dipole resonance (GDR)
in $^{16}$O.
Then, we compare the result with that of the former 1D continuum RPA
(which neglected the residual spin-orbit, Coulomb, and spin-spin parts),
discussing effects of the residual interactions.

We adopt the Skyrme energy functional as same as Eqs.~(A.2), (A.15), and (A.16)
in Ref.~\cite{BFH87}.
The static HF+BCS code based on this functional is called EV8
which assumes the parity and the $z$-signature symmetry.
In the present work, we do not assume any symmetry, in order to allow
a time-dependent state to be any Slater determinant 
during the time evolution.
The energy density is written in terms of local densities as
\begin{equation}
\label{energy_density}
{\cal H}(\vecr)=\frac{1}{2m}\tau(\vecr)
+{\cal H}^{\rm even}(\vecr)+{\cal H}^{\rm odd}(\vecr) ,
\end{equation}
with
\begin{eqnarray}
\label{H_even}
{\cal H}^{\rm even}(\vecr)&=&
 {\cal H}^{\rm even}[\rho^n,\rho\triangle\rho,\rho\tau,
  \rho\overleftrightarrow{\nabla}\cdot\overleftrightarrow{\mathbf{J}}], \\
\label{H_odd}
{\cal H}^{\rm odd}(\vecr)&=&
{\cal H}^{\rm odd}[\mathbf{j}^2,\mathbf{s}^2,
                   \mathbf{s}\cdot\nabla\times\mathbf{j}] .
\end{eqnarray}
Here, we follow the notation in Ref.~\cite{DD95}.
According to Ref.~\cite{BFH87},
terms of $\mathbf{s}\cdot\mathbf{T}-\overleftrightarrow{\mathbf{J}}^2$,
$\mathbf{s}\cdot\triangle\mathbf{s}$, and $(\nabla\cdot\mathbf{s})^2$
are omitted.
The energy functional, ${\cal H}^{\rm even}+{\cal H}^{\rm odd}$,
keeps the local gauge invariance \cite{Eng75,DD95}.
It is customary in the static HF calculation to take account of
the center-of-mass correction by multiplying the first term in
Eq.~(\ref{energy_density}) by a factor $(A-1)/A$.
We use this correction both for static and dynamic calculations.

In order to see effects of time-odd components,
we adopt the SIII interaction
which was used in the 1D continuum RPA calculation in Ref.~\cite{LG76}.
We perform the TDHF calculation with the full functional of
${\cal H}^{\rm even}+{\cal H}^{\rm odd}$,
and the one neglecting ${\cal H}^{\rm odd}$.
Hereafter, let us call the former functional ``SIII-full'',
and the latter ``SIII-even''.
The instantaneous external field is chosen as
\begin{equation}
\label{E1_field}
V_{\rm ext}(\vecr,t)=\epsilon {\cal M}(E1,\mu=0)\delta(t)
    =\epsilon e^{(E1)} rY_{10}(\hat{\vecr}) \delta(t),
\end{equation}
where $e^{(E1)}$ indicates the $E1$ recoil charge,
$Ne/A$ for protons and $-Ze/A$ for neutrons.
$\epsilon$ is an arbitrary small number.
Then, solving the TDHF equation
\begin{equation}
i\frac{\partial}{\partial t}\phi_i(t)=
\left\{-\frac{1}{2m}\nabla^2 + V^{\rm even}(t) + V^{\rm odd}(t)
\right\} \phi_i(t) ,
\end{equation}
for $i=1,\cdots,A$.
The time evolution is performed up to $t=30$ MeV$^{-1}$.
The time-even mean field, $V^{\rm even}$,
has been well tested against a large number of
experimental observations.
In order to test the time-odd mean field, $V^{\rm odd}$,
we need to investigate dynamical properties of nuclei.

Figure~\ref{fig: O16_GDR}~(a) shows time evolution of
calculated $E1$ dipole moment,
$\bra{\Psi(t)}\mathcal{M}(E1)\ket{\Psi(t)}$.
In the SIII-even calculation,
we see a beating pattern which results in two main peaks of
the dashed line in 
Fig.~\ref{fig: O16_GDR}~(b).
This is in a good agreement with the result of the 1D continuum RPA \cite{LG76}.
However, the inclusion of the time-odd mean field,
which is necessary for the Galilean invariance,
changes the strength distribution into a single peak (solid line).
We decompose effects of time-odd densities into those of 
current density $\mathbf{j}$ and
spin density $\mathbf{s}$
in Fig.~\ref{fig: O16_GDR}~(c) and (d), respectively.
The current density provides additional residual interaction to push
the GDR to higher energy by $0.5-1.3$ MeV,
while the spin density merges the two main peaks into one.
This effect of time-odd density is not special to the SIII interaction.
The same effect is observed with the SGII parameters of the Skyrme
interaction.
See Ref.~\cite{NY04-P2} for a brief report on the same calculation
with the SGII force.
It is somewhat surprising that not only the current but also the spin density
significantly modify the GDR structure.
Photoneutron cross section data \cite{BF75} are shown in the panel (c)
by a dotted line.
Their absolute values are multiplied by 3.5,
since the data indicates less than 20 \% of the TRK sum rule.
The experimental shape of the GDR resembles that of the SIII-even
calculation, but the two main peaks are calculated lower by about 3 MeV.
Agreement on the main peak position is slightly improved
in the SIII-full calculation,
though the calculated peak is still lower than the experiment by $2-2.5$ MeV.

If the interaction commutes with the $E1$ operator,
the oscillator sum
\begin{eqnarray}
S(E1)&\equiv& \int_0^\infty E B(E1; 0^+\rightarrow E(1^-)) dE, \\
&=& \sum_{\mu=-1}^1 \int_0^\infty 
     E \left|\bra{E(1\mu)} \mathcal{M}(E1,\mu) \ket{0}\right|^2 dE,
\end{eqnarray}
is identical to the TRK classical sum rule value
\begin{equation}
S(E1)_{\rm class}=\frac{9e^2}{8\pi m} \frac{NZ}{A} .
\end{equation}
For $^{16}$O, the classical sum rule gives
$S(E1)_{\rm class}=59.4$ e$^2$ fm$^2$ MeV.
Since the Skyrme interaction has momentum and isospin dependence,
the classical sum rule is violated to a certain extent.
We have $S(E1)=75.1$ e$^2$ fm$^2$ MeV for SIII-full, and
$S(E1)=67.1$ e$^2$ fm$^2$ MeV for SIII-even.
The enhancement of the TRK sum is 26 \% for SIII-full
and it reduces to 13 \% if we neglect the time-odd mean field.
This difference mainly comes from the spin density.
If we integrate the strength in the energy region up to 30 MeV,
we have $S(E1)\approx S(E1)_{\rm class}$ for both SIII-full and SIII-even.

\subsection{$E1$ resonances in even-even Be isotopes}
\label{sec: Be}

Finally, we apply the small-amplitude TDHF+ABC method to $E1$ resonances in
beryllium isotopes.
Beryllium nuclei have been extensively studied both theoretically and
experimentally (see, e.g., Ref.~\cite{Cluster04} and references therein).
$^8$Be is well-known for the $\alpha$-$\alpha$ clustering structure
with an elongated prolate shape.
Valence neutrons added to $^8$Be are expected to cause variety of
structure change in the ground and excited states \cite{SKN81,KHO95,IOI00}.
$^{10}$Be has two neutrons in addition to $^8$Be.
The $\alpha$-$\alpha$ distance in the ground state
is considered to be slightly smaller than that in $^8$Be.
$^{12}$Be is a semi-magic nucleus ($N=8$), however
its properties are different from spherical closed-shell nuclei.
The measured spectroscopic factors suggest that the last neutron pair is
two-thirds in the $sd$ configurations \cite{Nav00}.
The neighboring odd nucleus, $^{11}$Be, is famous for the parity inversion
and for the halo structure in the ground state.
The existence of $^{14}$Be at the drip line beyond $N=8$ also indicates
weakening of shell closure at $N=8$.
Since both $Z=4$ and $N=10$ are the magic numbers at 
the prolate superdeformed shape,
we expect $^{14}$Be to be deformed as large as $^8$Be.
A new mode of excitation of significant interest
is the soft $E1$ mode near the neutron drip line \cite{HJ87,Ike92}.
Coupling in the continuum between the soft $E1$ mode and
the quadrupole deformation
is an unsolved problem which can be addressed by the present method
to some extent.

We use the Skyrme interaction of the SIII parameters including the time-odd
components (SIII-full).
The adopted model space is the adaptive grid in Fig.~\ref{fig: grid}
with $R=10$ fm and $\varDelta r =12$ fm.
Usually, the static HF calculation is carried out with constraint on
the center of mass at the origin.
However, in this calculation, we do not impose any condition
on the center-of-mass and on the direction of the principal axis.
Although this results in heavy computation for the imaginary-time step,
it turns out that this is important for the stable time evolution of
the TDHF state kicked off by the external perturbation.
The external field is the same form as Eq.~(\ref{E1_field}), but
includes $rY_{1\pm 1}$.
The TDHF calculation with the perturbative $E1$ field
provides the $E1$ intrinsic strength, $dB(\omega, {\cal M}(E1,K))/d\omega$
through Eq. (\ref{BF_RT}).
Assuming the strong coupling scheme \cite{BM75},
the $B(E1)$ transition strength in the laboratory frame is given by
\begin{eqnarray}
\label{BE1_1}
\frac{dB(\omega; E1)}{d\omega} &\equiv&
\int dE_x\ B(E1; 0^+\rightarrow E_x(1^-)) \delta(\omega-E_x),
 \nonumber \\
\label{BE1_2}
&=&\sum_{K}\int dE_x |\bra{E_x} {\cal M}(E1,K) \ket{0}|^2
   \delta(\omega-E_x) , \nonumber \\
\label{BE1_3}
&=&\sum_{K=0,\pm 1} \frac{dB(\omega; {\cal M}(E1,K))}{d\omega} .
\end{eqnarray}
Here, the state $\ket{0}$ ($\ket{E_x}$) is the intrinsic ground (excited)
state.

The static HF calculation predicts all these nuclei to be deformed
in prolate shape in the ground state.
Calculated quadrupole deformations are given in Table~\ref{tab: Be_def}.
As we expected, $^8$Be and $^{14}$Be possess large deformation.
$^{12}$Be has the smallest deformation, but its proton distribution
has a moderate deformation.
The static HF analysis on Be isotopes with the SIII force
have been already done in Ref.~\cite{TYI95}.
The total binding energies are well reproduced.
Calculated occupied single-particle energies are
listed in Table~\ref{tab: Be_spe}.
In the linear response approximation of the TDHF,
the neutron continuum plays its role at energies higher than
the absolute value of single-particle energy of
the last-occupied neutron.
Since the proton orbitals become deeply bound in neutron-rich nuclei,
the proton continuum is expected to be less important.
However, in these Be isotopes, the protons are important to produce
prolate deformation of the mean field.

\begin{table}
\caption{Calculated quadrupole deformation for even-even Be isotopes.
The last two columns show deformation of neutron and proton density
distribution, $\beta_n$ and $\beta_p$, separately.}
\begin{ruledtabular}
\begin{tabular}{r|lll}
          & $\beta$ & $\beta_n$ & $\beta_p$ \\
\hline
$^8$Be    & 1.07 & 1.06 &  1.09 \\
$^{10}$Be & 0.39 & 0.33 &  0.49 \\
$^{12}$Be & 0.12 & 0.07 &  0.22 \\
$^{14}$Be & 0.74 & 0.77 &  0.66 \\
\end{tabular}
\end{ruledtabular}
\label{tab: Be_def}
\end{table}
\begin{table}
\caption{Calculated neutron (n) and proton (p)
single-particle energies in units of MeV for Be isotopes.
Each state has a two-fold degeneracy associated with
the time-reversal symmetry.}
\begin{ruledtabular}
\begin{tabular}{rr|rr|rr|rr}
\multicolumn{2}{c|}{$^8$Be} &
\multicolumn{2}{c|}{$^{10}$Be} &
\multicolumn{2}{c|}{$^{12}$Be} &
\multicolumn{2}{c}{$^{14}$Be} \\
\multicolumn{1}{c}{n} & \multicolumn{1}{c|}{p} &
\multicolumn{1}{c}{n} & \multicolumn{1}{c|}{p} &
\multicolumn{1}{c}{n} & \multicolumn{1}{c|}{p} &
\multicolumn{1}{c}{n} & \multicolumn{1}{c}{p} \\
\hline
 -24.8 &  -22.9 & -26.3 &  -31.6 & -25.5 &  -36.7 & -24.9 &  -39.1 \\
 -13.3 &  -11.5 & -12.5 &  -16.0 & -11.5 &  -20.6 & -14.0 &  -25.7 \\
       &        &  -9.7 &        & -10.3 &        &  -9.1 & \\
       &        &       &        &  -4.3 &        &  -3.5 & \\
       &        &       &        &       &        &  -2.6 & \\
\end{tabular}
\end{ruledtabular}
\label{tab: Be_spe}
\end{table}

\begin{figure}[ht]
\includegraphics[width=0.45\textwidth]{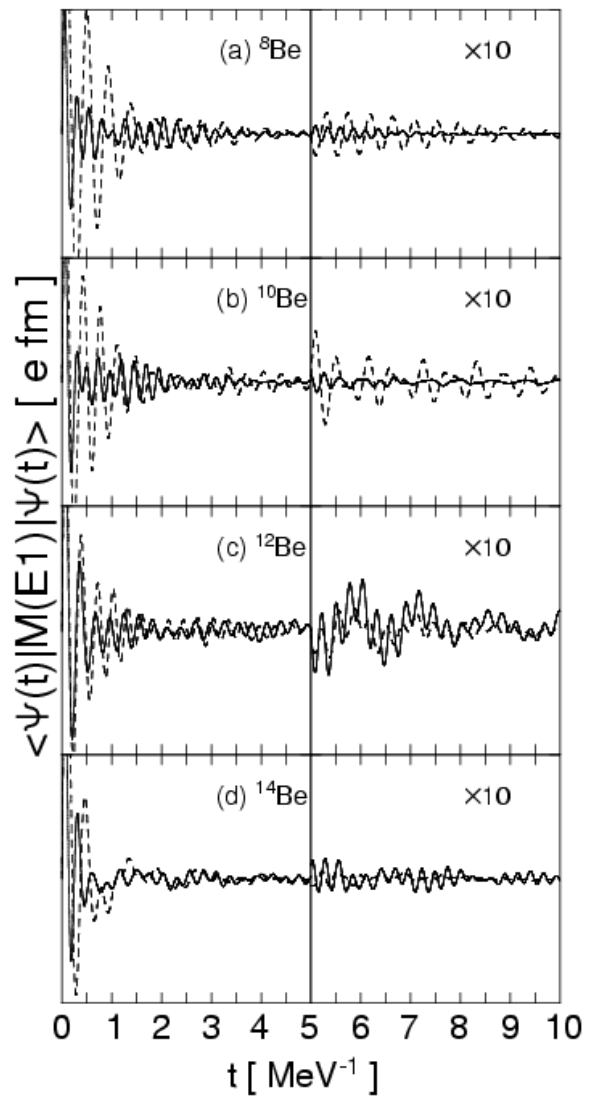}
\caption{\label{fig: Be_RT}
Calculated $E1$ moment as functions of time
for (a) $^8$Be, (b) $^{10}$Be,
(c) $^{12}$Be, and (d) $^{14}$Be.
The dashed (solid) line indicates the external field with $K=0$ ($K=\pm 1$).
Scale of the vertical axis is arbitrary because it linearly depends
on the small parameter $\epsilon$.
It is magnified by a factor of five
for the latter half of period, $5<t<10$ MeV$^{-1}$.
}
\end{figure}

\begin{figure}[ht]
\includegraphics[width=0.45\textwidth]{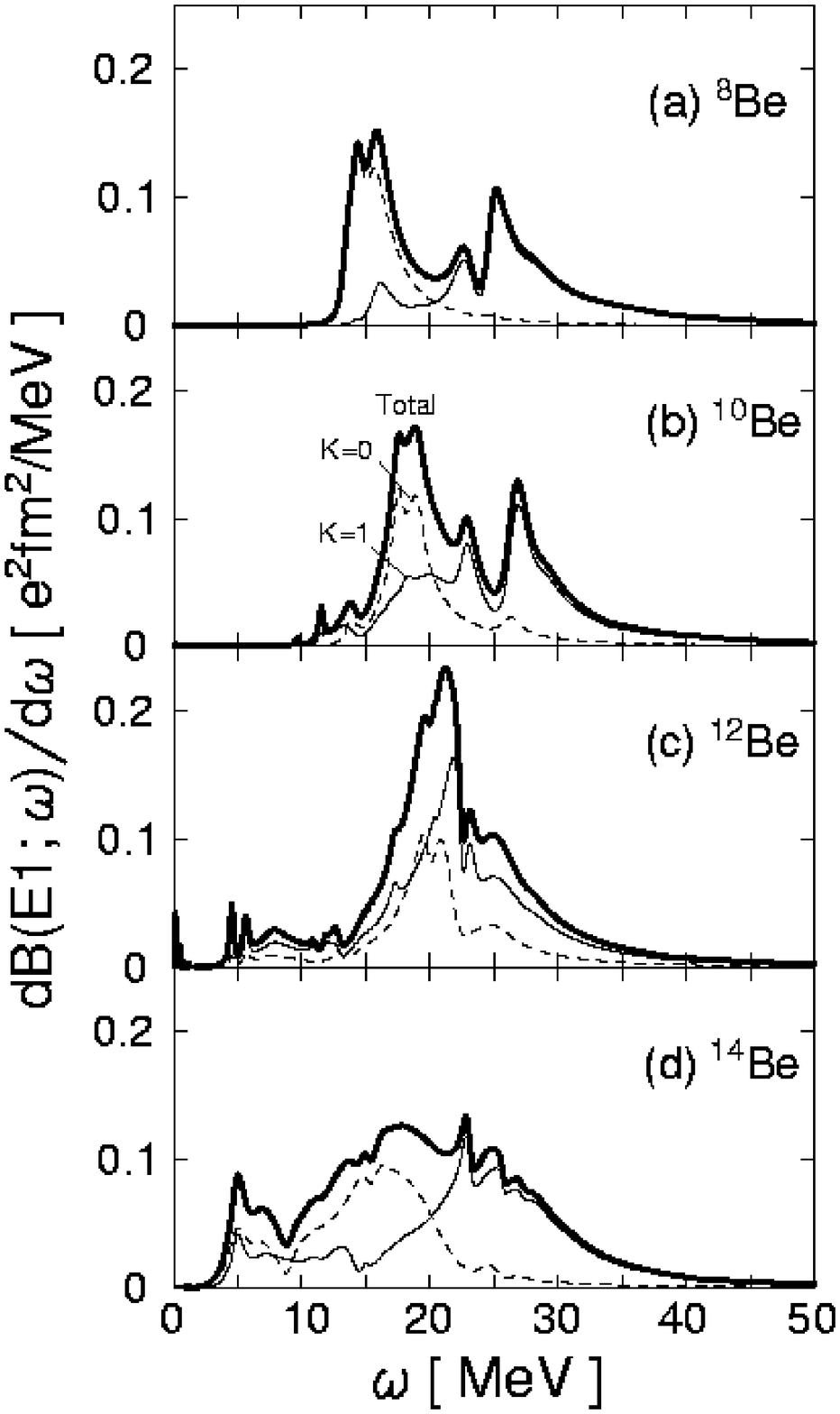}
\caption{\label{fig: Be_GDR}
Calculated values of
$dB(E1; 0\rightarrow 1^-)/d\omega$
for $^{8,10,12,14}$Be.
The smoothing parameter $\Gamma=0.2$ MeV is used.
The thin dashed (solid) line is a contribution of $K^\pi=0^-$ $(1^-)$
states and the thick solid line for the total strength.
}
\end{figure}


Now let us discuss dynamical properties of these nuclei.
In Fig.~\ref{fig: Be_RT}, calculated time-dependent $E1$ moment 
is presented as functions of time.
The time evolution is calculated up to $t=30$ MeV$^{-1}$.
The beginning third of the total period is shown in Fig.~\ref{fig: Be_RT}.
Note that the amplitude is magnified by a factor of 10 in the latter half
of period.
Performing the Fourier transform,
calculated $B(E1)$ transition strength is shown in Fig.~\ref{fig: Be_GDR}.
In $^8$Be, we observe large splitting of the GDR peak associated with
the large quadrupole deformation ($\beta\approx 1$).
The magnitude of splitting is more than 10 MeV,
that brings down the $K=0$ peak (the oscillation along the symmetry axis)
to around 15 MeV in energy.
In Fig.~\ref{fig: Be_RT} (a), 
the amplitude of the $K=0$ oscillation almost monotonically decays
as time increases, while that of the $K=1$ mode shows a beating pattern.
This results in a splitting of the high-lying $K=1$ peak.
In the total $B(E1)$ strength function,
this is seen as a small peak in the middle of two main peaks.
We also see that the $K=0$ oscillation stays longer than the $K=1$.
This is because the peak position is near the particle decay
threshold, thus, the allowed phase space is smaller for
the $K=0$ peak (Table~\ref{tab: Be_spe}).

In $^{10}$Be, we see a similar behavior to $^8$Be.
Because of the smaller deformation, the lower $K=0$ peak shifts to
higher energy by about 5 MeV.
Figure \ref{fig: Be_GDR} (b) again indicates the $K=1$ mode split
into two peaks.
Although the ground-state deformation in $^{10}$Be is less than half of $^8$Be,
the energy splitting between the lowest and highest peaks
is still as large as $7\sim 8$ MeV.

Next, let us discuss $^{12}$Be.
The calculated quadrupole deformation is the smallest among
these even-even isotopes.
In contrast to $^{8,10}$Be,
we do not see a distinguished double-peak structure.
The GDR shows a peak at $21$ MeV with a broader structure around 25 MeV.
There exist low-energy $E1$ strength in the continuum below 10 MeV.
The peak very near zero energy is due to small admixture of the
translational mode.
Because of this mixing,
the response function suffers from spurious oscillatory behavior
at energy below 2 MeV.
Thus, we concentrate our focus on states at $\omega>2$ MeV.

Though the $B(E1)$ strength in low energy looks small
compared to that in the main GDR,
the integrated strength in the energy region of
$2<\omega<10$ MeV amounts to $B(E1)\approx 0.14$ e$^2$ fm$^2$.
The lowest sharp peak at 4.5 MeV has
$B(E1; 0^+\rightarrow 1^-)\approx 0.023$ e$^2$ fm$^2$.
The next lowest peak at 5.6 MeV has $B(E1)\approx 0.027$ e$^2$ fm$^2$.
Both peaks have a dominant $K=1$ character.
The low-lying $1^-$ state has been recently observed in $^{12}$Be \cite{Iwa00}.
The observed excitation energy is $E_x=2.68(3)$ MeV with
$B(E1; 0^+\rightarrow 1^-)= 0.051(13)$ e$^2$ fm$^2$.
Our result is higher in energy by a factor of two
and the sum of $B(E1)$ for the lowest two peaks is comparable to
the experiment.
The two-neutron pairing model in Ref.~\cite{BM97} predicted the $1^-$ energy
very well (2.7 MeV) but about five times overestimated $B(E1)$.
The shell model calculation with extended single-particle wave functions
in Ref.~\cite{SSII01} well reproduced the lowest $1^-$ state
($E_x=2.14\sim 2.9$ MeV with $B(E1)=0.063\sim 0.072$ e$^2$ fm$^2$
depending on the interaction and model space).
They also calculated $B(E1)$ strength distribution in the GDR energy region
without taking account of the continuum.
Although their results strongly depend on the adopted interaction and
model space, the calculated GDR energy is lower than ours.
A striking difference from our result is that they have predicted three
main peaks with the WBP interaction.
It is not clear at present whether this difference is due to the treatment
of the continuum or to the ground-state correlation.

Finally, let us move to the drip line, $^{14}$Be.
The doubly-magic closed-shell configuration ($N=10$, $Z=4$)
at superdeformation leads to the large quadrupole deformation
of $\beta=0.74$.
The $K=0$ and $K=1$ resonance peaks are at different positions
whose centroids are at 15 MeV and 24 MeV.
Figure \ref{fig: Be_RT} (d) indicates quick damping of the
$K=0$ oscillation.
The oscillating pattern almost disappears by $t=3$ MeV$^{-1}$.
This leads to the large width of the $K=0$ peak in Fig.~\ref{fig: Be_GDR} (d).
As a consequence of the large width, 
the double-peak structure in the total $B(E1)$ strength function
is not as clear as in $^{8,10}$Be.
It looks more like a single broad resonance at 20 MeV with the width of
about 20 MeV.
In Fig. \ref{fig: Be_RT} (d),
after the $K=0$ mode disappears,
the $K=1$ mode becomes dominant at $t>3$ MeV$^{-1}$.
This long-lived high-frequency $K=1$ mode results in
sub-peaks embedded in the broad $K=1$ resonance ($20<\omega<25$ MeV).

It is known that the weakly bound neutrons strongly couples to the
continuum and produces the large dipole strength \cite{BW52}.
The Coulomb breakup of $^{11}$Be is a typical example \cite{Nak94}.
This is often called ``threshold effect'' which has a peak at the
threshold energy.
Since the SIII interaction gives the last neutron binding of $2-3$ MeV,
the threshold effect is weak.
In the present calculation, we do not have significant threshold strength.
On the other hand, another soft dipole peak is seen at 5 MeV.
This peak carries $B(E1)\approx 0.26$ e$^2$ fm$^2$.
A Coulomb dissociation experiment seems to suggest enhanced strength
at $E_x\approx 2$ and 5 MeV \cite{Lab01}.
The shell-model calculation of Ref.~\cite{SSII01} also indicates a similar
peak ($E_x=6.76\sim 7.46$ MeV with $B(E1)=0.097\sim 0.146$ e$^2$ fm$^2$).
Using the SGII interaction, this peak is at 7 MeV with
$B(E1)\approx 0.14$ e$^2$ fm$^2$ \cite{NUY04-P}, which well agrees with
the shell-model result.
On the GDR main peaks, our result looks rather different from
the shell-model:
the shell-model indicates a single  main
peak at $12-17$ MeV, while we have a broad resonance whose centroid
is around 20 MeV.
Since the shell model also indicates continuous $E1$ strength in the 
energy region of $\omega>10$ MeV,
this difference may be simply due to lack of the continuum
in Ref.~\cite{SSII01}.

Calculated TRK sum rule values are listed in Table~\ref{tab: Be_sum_rule}.
The enhancement is slightly smaller than the spherical $^{16}$O case.
Among the even-even Be isotopes, the enhancement is the biggest for $^{12}$Be
whose deformation is the smallest.
According to the analysis on $^{16}$O in Sec.~\ref{sec: O16},
about the half of this enhancement comes form effect of the time-odd
spin density.
The large deformation leads to a strong coupling to the $K$-quantum number,
and this may restrict dynamics of spin degrees of freedom.
The soft $E1$ strength,
which is defined by the oscillator sum up to 15 MeV in the table,
varies among these isotopes.
The large value in $^8$Be is due to the large ground-state deformation
that brings the low-energy $K=0$ peak down close to 14 MeV.
In $^{14}$Be, it is the largest.
This is due to combination of the deformation, the soft dipole peak at 5 MeV,
and the large width of the $K=0$ resonance at 15 MeV.
The deformation parameter $\delta$ is estimated from the average energies
of $K=0$ and $K=1$ modes.
We use Eq. (6-344) in Ref.~\cite{BM75}.
The $\delta$ turns out to be much smaller than the deformation of the HF
density distribution, $\beta$.
The deformation derived from the GDR splitting is known to well agree with
that from the $E2$ moment for actinide nuclei \cite{BM75}.
In light nuclei, the geometrical interpretation of the GDR frequencies
may not be justified so well.

\begin{table}
\caption{
Energy weighted sum rule values in units of e$^2$ fm$^2$ MeV.
The second column shows values of the classical TRK formula.
The small-amplitude TDHF+ABC calculation produces values in the third columns.
The fourth column gives the soft $E1$ strength of
the energy weighted sum, which is defined by $\omega<15$ MeV.
The last column indicates the deformation parameter obtained by the
splitting of the GDR peaks.
}
\begin{ruledtabular}
\begin{tabular}{rdddd}
        &  \multicolumn{1}{c}{$S(E1)_{\rm class}$} &
           \multicolumn{1}{c}{$S(E1)$} &
           \multicolumn{1}{c}{$S(E1; E<15\mbox{MeV})$} &
           \multicolumn{1}{c}{$\delta$} \\
\hline
{$^8$Be}    & 29.7 & 34.0 & 3.14 & 0.43 \\
{$^{10}$Be} & 35.7 & 42.8 & 1.26 & 0.21 \\
{$^{12}$Be} & 39.6 & 48.2 & 2.54 & 0.05 \\
{$^{14}$Be} & 42.5 & 52.2 & 7.57 & 0.35 \\
\end{tabular}
\end{ruledtabular}
\label{tab: Be_sum_rule}
\end{table}

\section{Conclusion}
\label{sec: Conclusion}

We have developed the linear response theory in the continuum
applicable to deformed systems.
The exact treatment of the continuum is done by the iterative method
for construction of the Green's function in the 3D Cartesian grid space
(3D continuum RPA).
The method is identical to the conventional 1D continuum RPA in
the spherical limit.
At the same time,
we have shown that the approximate but yet accurate treatment of the continuum
can be done by the absorbing boundary condition (ABC) approach.
The small-amplitude TDHF+ABC method in the linear response regime
is practically identical to the 3D continuum RPA.
Applications of these methods to the TDHF with the BKN interaction
reveals their usefulness and accuracy.
The real-time TDHF method has a difficulty when we study
excitation modes coupled to the zero modes.
Since the method is fully self-consistent,
the increase of model space (finer grid spacing) will solve the problem,
though it requires heavier computation.

Applications to systems with a realistic effective interaction have been
performed with the small-amplitude Skyrme TDHF+ABC.
The analysis on the GDR in $^{16}$O suggests a significant contribution coming
from the time-odd mean field which was often neglected in the 1D continuum RPA.
The peak structure in the continuum is affected by these residual interactions,
especially by the spin density.
Since the spin-dependent terms in the Skyrme energy functional,
such as $\mathbf{s}^2$,
$\mathbf{s}\cdot\triangle\mathbf{s}$, and $(\nabla\cdot\mathbf{s})^2$,
are not linked to the time-even components by the local gauge invariance,
the analysis may give a useful constraint on these parts of
the Skyrme functional.

The coupling to the continuum becomes more important for weakly
bound systems.
We have studied the deformed continuum of the GDR in Be isotopes.
The large deformation splitting of about 10 MeV is predicted for $^{8,14}$Be.
The $K=0$ main peak is significantly lowered by the deformation
to less than 15 MeV.
The time evolution of the $E1$ moment indicates different damping
between $^8$Be and neutron-rich Be isotopes,
especially for the $K=0$ dipole mode.
The soft dipole strength ($E<10$ MeV) appears in $^{12}$Be and $^{14}$Be.
Considering the fact that the SIII parameters were not determined by
the isovector properties,
we have a reasonable agreement with experiment
on the low-energy $1^-$ state in $^{12}$Be and $^{14}$Be.

In this paper, we have studied only the IV GDR in neutron-rich nuclei,
because of the numerical difficulty discussed above.
The IS modes in neutron-rich deformed nuclei are
also interesting to investigate.
For instance, the octupole correlation in superdeformed $^{14}$Be is
expected to be stronger than $^8$Be.
This is because the superdeformed magic numbers are classified into
two category, and
the $N=10$ shell closure has a stronger octupole correlation
than the $N=4$ \cite{ND92,NMM92}.
The small-amplitude TDHF+ABC may be a good method
to see how the continuum affects this expectation,

An important extension of the present approaches is the inclusion of pairing.
Since the pairing plays an important role in heavy nuclei,
this is very desirable but a difficult task.
In this respect, we should mention that
the HFB-based continuum QRPA has been recently proposed by
Matsuo \cite{Mat01} to take account of the continuum for both particle-hole
(p-h) and particle-particle (p-p)/hole-hole (h-h) channels.
The combination of the 3D continuum RPA and
the continuum QRPA may produce a general theory to calculate
excited states in the p-h, p-p, and h-h continuum for nuclei
in the whole nuclear chart.

\begin{acknowledgments}
This work has been supported by the Grant-in-Aid for Scientific Research
in Japan (Nos. 14540369 and 14740146), and
done as a part of the Japan-U.S. Cooperative
Science Program ``Mean-field approach to collective excitations in unstable
medium-mass and heavy nuclei".
A part of the numerical calculations
have been performed on the supercomputer at the Research Center for
Nuclear Study (RCNP), Osaka University.
\end{acknowledgments}

\bibliography{myself,nuclear_physics,chemical_physics}

\end{document}